\documentstyle[12pt]{article}
\begin{document}

\author{Pawel O. Mazur 
\thanks{Electronic address: mazur@swiatowid.psc.sc.edu}
\\
\small{\it  Department of Physics and Astronomy,}
\\
\small{\it University of South Carolina,}
\\
\small{\it Columbia, S.C. 29208}}

\title{Reply to Comment on 
``Spinning Cosmic Strings and Quantization of Energy''}

\date{\small (24 November 1996)}
\maketitle

The preceding comment \cite{Sam} has raised the question of the validity of 
the argument presented in the Letter \cite{Maz1} that the energy of the 
particles should be quantized in the background of a spinning cosmic string. 
As the construction of the electromagnetic analogy given in \cite{Maz1} 
suggests, a spinning cosmic string is the gravitational analog of the 
Aharonov-Bohm solenoid (flux tube), and the angular momentum per unit 
length $J$ 
corresponds to the magnetic flux $\Phi $ and the mass per unit length $m$ 
corresponds to the charge per unit length\footnote[1]{Strictly speaking, it 
corresponds to the charge; added on November 24, 1996.} $q$. Mathematically, 
the problem of the quantum field propagating in the background of a 
spinning cosmic string is analogous to the quantum mechanics of charged 
particles in the field of the Aharonov-Bohm solenoid \cite{Aha}. One expects, 
therefore, the appearance of the gravitational Aharonov-Bohm effect. In fact, 
it can be easily shown that the gravitational scattering cross section of 
particles with energy $E$ in the background of a spinning string with angular 
momentum $J$ and mass $m$ per unit length is \cite{Maz2} \footnote[2]{This 
paper has never been submitted for publication; the results of this work were 
described in several seminars in the fall of 1986 and in the spring of 1987, 
and it was not quite understood by my audience; added on November 24, 1996.} 
\begin{eqnarray*}
\frac{d^{2}\sigma }{dzd\vartheta } &=&\hbar \frac{\sin ^{2}(\pi \beta
/\alpha )}{2\pi E}\left[ \sin ^{2}\left( \frac{\pi }{2\alpha }-\frac{%
\vartheta }{2}\right) \right] ^{-1}, \\
\beta  &=&4GJE/\hbar ,\alpha =1-4Gm.
\end{eqnarray*}
The spinning string is ``transparent'', i.e., it cannot be detected by
scattering experiments, only if the energy $E$ of particles satisfies the 
quantization condition obtained in \cite{Maz1}, $E=\hbar n\alpha /4GJ.$ 

One may adopt two different points of view about this quantization condition; 
one of them is presented in \cite{Maz1}. Namely, we observe that the external 
metric of the idealized infinitely thin spinning string has the 
causality-violating region, which one might say is physically unacceptable 
if it is not shielded by the event horizon or if it is not inside the matter 
(i.e., if it is not the case that the thickness of a spinning string 
is greater than the size of the causality-violating region). 
The presence of causality-violating regions 
would be detected unless particles have quantized energies for which a 
spinning string is ``transparent''. This quantization condition on the 
energy of particles propagating in the background of a spinning string is 
mathematically equivalent to the imposition of boundary conditions periodic 
in time on the wave functions. Effectively the space-time of the spinning 
string is periodic in time. 

There is another possible interpretation of the above-mentioned result. 
Namely, the spinning string is the real thing and it has some finite 
thickness, which, in general, might be greater than the size of the the 
causality-violating region. The external metric of this string will still be 
described by Eq. (3) of \cite{Maz1}. 
In such a case a spinning string will cause 
the characteristic Aharonov-Bohm-type scattering of particles. It is the 
physically sound requirement of nondetectability of causality-violating 
regions which leads us to postulate the quantization condition which, on the 
other hand, is not logically implied by physics of string.

{\it Note added on November 24, 1996}. 

After this note was written \footnote[3]{This short note was written 
at the end of November 1986 when the author was a research associate 
at Syracuse University; so, it was never 
distributed in the preprint form.}, approximately 2 years later, 
several papers appeared which contain the same idea and describe 
the similar calculation [7, 8, 9]. 
Once established, the Aharonov-Bohm character of quantum mechanical 
interaction between (spinning) particles in $2+1$-dimensional gravitation 
of Staruszkiewicz [5, 6] has led immediately to the gauge model reformulation 
\cite{Witt}, which is physically rather misleading . The latter work 
also extends some earlier work on the abelian gauge model of the same type 
\cite{Polyak}.   
This author presented the basic idea 
of the new quantization condition for the gravitational mass-energy, which 
involves also the Newton constant $G$ (in addition to the Planck constant $h$) 
in the paper published some time later \cite{Maz3}. 
The quantized spectrum of energy for the gravitational two-body problems 
of massless particles interacting with the spinning string
\footnote[4]{The spinning string of [2, 14] is a torsion vortex 
and should be considered an analog of the Abrikosov 
vortex in the context of the Palatini formulation 
of Einstein's General Relativity Theory.} and the massless particle in the 
s-wave sector interacting with a Schwarzschild black hole must follow from the 
same physical principle \cite{Maz3}. 
It was then suggested that quantization of the 
gravitational mass-energy, and the existence of the gravitational quanta 
(the space-time-matter quanta) must follow from some fundamental physical 
principle such as the Universal Second Law \cite{Maz4}.


\begin{thebibliography}{99}
\bibitem{Sam}  J. Samuel and B. R. Iyer, preceding Comment [Phys. Rev. Lett. 
\textbf{59}, 2379 (1987)].
\bibitem{Maz1}  P. O. Mazur, Phys. Rev. Lett. \textbf{57}, 929 (1986).
\bibitem{Aha}  Y. Aharonov and D. Bohm, Phys. Rev. \textbf{115}, 485 (1959).
\bibitem{Maz2}  P. O. Mazur, to be published.

\textbf{References added on November 24, 1996:}

\bibitem{Star}  A. Staruszkiewicz, Acta Phys. Polon. \textbf{24}, 734 (1963).
\bibitem{Th1}  S. Deser, R. Jackiw, and G. t' Hooft, Ann. Phys. \textbf{152}, 
220 (1984).
\bibitem{Th2}  G. 't Hooft, Commun. Math. Phys. \textbf{117}, 685 (1988).
\bibitem{DJ}  S. Deser and R. Jackiw, Commun. Math. Phys. \textbf{118}, 495 
(1988).
\bibitem{dSJack}  P. de Sousa and R. Jackiw, Commun. Math. Phys. 
\textbf{124}, 229 (1989).
\bibitem{Witt}  E. Witten, Nucl. Phys. \textbf{B311}, 46 (1988).
\bibitem{Polyak}  A. M. Polyakov, Mod. Phys. Lett. \textbf{A3}, 325  (1988).
\bibitem{Maz3}  P. O. Mazur, Acta Phys. Polon. \textbf{B26}, 1685 (1995).
\bibitem{Maz4}  P. O. Mazur, Acta Phys. Polon. \textbf{B27}, 1849 (1996).
\bibitem{KMP}  K. Koltko, P. O. Mazur, and W. Puszkarz, 
submitted to Phys. Rev. \textbf{D}.

\end{thebibliography}
\end{document}